\documentclass[11pt,centertags,reqno,twoside]{amsart}
\usepackage{amsfonts}
\usepackage{amsmath}
\usepackage{amssymb}
\usepackage{mathptmx}
\usepackage[dvips]{graphicx}
\usepackage{epsfig}

\begin{document}

\title[Faddeev-Yakubovsky Differential Equations: Methods]
{Progress in Methods to Solve the
Faddeev and Yakubovsky Differential Equations$^*$}

\author[A.\,K.\,Motovilov]
{Alexander K. Motovilov}

\address{Bogoliubov Laboratory of Theoretical Physics,
JINR, Joliot-Curie 6, 141980 Dubna, Russia}

\thanks{\noindent$^*$Based on a talk given at the
\htmladdnormallink{20th European Conference on Few-Body Problems in
Physics}{http://www.pi.infn.it/efb20/} (September 10--14, 2007,
Pisa, Italy). The work was supported by the Deutsche
Forschungsgemeinschaft (DFG) and the Russian Foundation for Basic
Research.}

\date{December 4, 2007}



\begin{abstract}
We shortly recall the derivation of the Faddeev-Yakubovsky
differential equations and point out their main advantages. Then we
give a review of the numerical approaches used to solve the
bound-state and scattering problems for the three- and four-body
systems based on these equations. A particular attention is payed to
the latest developments.
\end{abstract}

\maketitle

\section{Brief history}

A crucial step in the theory of few-body quantum systems has been
done in 1960 when Faddeev introduced his celebrated integral
equations \cite{Faddeev1960}. The Faddeev equations became a basis
for the three-body scattering theory with short-range interactions.
They have also been used through the years in numerical
calculations. As for the $N$-body systems with $N>3$, Faddeev's idea
of an explicit clus\-ter-chan\-nel separation was completely
elaborated by 1967 when Yakubovsky's integral equations became
available \cite{Yakubovsky1967}. The Faddeev differential equations,
in their $S$-wave two-dimensional version, were first
discovered in 1968 by Noyes and Fiedeldey \cite{NF1968}. The Faddeev
differential equations in their complete form have been analysed in
1976 by Merkuriev, Gignoux and Laverne \cite{MGL1976}. In
particular, they found and approved the asymptotic boundary
conditions that are needed to be added to the equations in order to
find unique physical solutions corresponding to various scattering
processes.

It was the great success of the Faddeev differential equations that
stimulated Mer\-ku\-ri\-ev to look for a generalization of these
equations to few-body systems with arbitrary number of particles. By
1982 he has successfully solved the problem jointly with his then
Ph.D. student Ser\-gei Ya\-kov\-lev \cite{MYa1982}. They not only
found the ``right'' differential equations for the Ya\-ku\-bov\-sky
components but also described the boundary-value problems for these
equations that correspond to certain scattering processes
\cite{MYa1983}. In a somewhat more abstract form, the
Ya\-ku\-bov\-sky differential equations can be seen in paper
\cite{BGH1982} by Benoist-Gueutal and L'Huillier, published also in
1982.

In Section 2 we recall how the Faddeev-Ya\-ku\-bov\-sky differential
equations look like. The best way to do this is simply to derive
them. We concentrate only on the algebraic context of the derivation,
leaving apart the description of the scattering boundary conditions.
In Section 3 we review the numerical approaches used to solve the
Faddeev-Ya\-ku\-bov\-sky differential equations. A particular
attention is payed to the latest developments. Finally, in Section 4
we mention some still challenging problems.

\section{Faddeev and Yakubovsky differential equations}

Let $H_0$ be a linear (not necessarily Hermitian) operator on a
Hilbert space $\mathfrak{H}$, and $V$ another linear operator on
$\mathfrak{H}$. At the moment we specify neither $H_0$, nor $V$.
These operators may be of completely arbitrary nature. Temporarily
assuming that both $H_0$ and $V$ are bounded we avoid complications
with domains of the operators involved and concentrate mainly on the
algebraic part of the Faddeev-Yakubovsky scheme. The only essential
assumption is that by some reason the perturbation $V$ is split into
the sum $V=V_1+V_2+\cdots+V_n$ of $n$ ($2\leq n<\infty$) terms
$V_\alpha$. Surely, one is interested in the perturbed operator
$H=H_0+V$.

Suppose that $z$ is an eigenvalue of $H$ and $\Psi$ the
corresponding eigenvector. Assume, in addition, that $z$ does not
belong to the spectrum $\sigma(H_0)$ of the unperturbed operator
$H_0$. Then $H_0-z$ is invertible and $(H_0+V)\Psi=z\Psi$ is
equivalent to the Lippmann-Schwinger equation
\begin{equation}
\label{Psi}
\Psi=-(H_0-z)^{-1}\sum\limits_{\alpha=1}^n V_\alpha\Psi.
\end{equation}
Introduce the vectors
\begin{equation}
\label{FC}
\psi_\alpha=-(H_0-z)^{-1}V_\alpha\Psi,\qquad \alpha=1,2,\ldots,n.
\end{equation}
These vectors are called the \textit{Faddeev components} of the
eigenvector $\Psi$. Obviously, combining \eqref{Psi} and \eqref{FC}
implies $ \sum\limits_{\beta=1}^n \psi_\beta=\Psi.$ Taking this
into account, rewrite \eqref{FC} in the form
\begin{equation}
\label{FE0}
\psi_\alpha=-(H_0-z)^{-1}V_\alpha\sum\limits_{\beta=1}^n \psi_\beta.
\end{equation}
Then apply to both sides of \eqref{FE0} the operator $H_0-z$ and
arrive at the equations
\begin{equation}
\label{FEc}
(H_0+V_\alpha-z)\psi_\alpha=
-V_\alpha\sum\limits_{\beta\neq\alpha}\psi_\beta.
\end{equation}
These are just the celebrated \textit{Faddeev ``differential''
equations} for the components $\psi_\alpha$ of an eigenvector of
$H$. A remark on the spectrum of the associated \textit{Faddeev operator}
can be found in Appendix.

For $z\not\in\sigma(H_0+V_\alpha)$, $\alpha=1,2,\ldots,n,$ one
can invert the operators \mbox{$H_0+V_\alpha-z$} and, thus, rewrite
\eqref{FEc} in the equivalent form
\begin{equation}
\label{FEint} \psi_\alpha=
-(H_0+V_\alpha-z)^{-1}V_\alpha\sum\limits_{\beta\neq\alpha}\psi_\beta.
\end{equation}
Equations \eqref{FEint} are known as the \textit{Faddeev integral
equations}.

The Faddeev equations \eqref{FEc} turn into truly differential
equations if $H_0$ is a differential operator as this happens in the
case of few-body problems with pairwise interactions, studied in
coordinate representation. In this case $H_0$ is simply the kinetic
energy operator in the center-of-mass frame and $V_\alpha$'s stand
for the two-body potentials.

In the three-body case the Faddeev equations \eqref{FEc} or
\eqref{FEint} represent simultaneously the first and the last step
in the Fad\-de\-ev-Ya\-ku\-bov\-sky approach. This reflects the fact
that with sufficiently rapidly decreasing and smooth potentials
$V_\alpha$, $\alpha=1,2,3,$ the four times iterated Fad\-de\-ev
integral equations \eqref{FEint} are Fredholm and even compact
(unlike the Lipp\-man-Schwin\-ger equation \eqref{Psi} that remains
non-Fred\-holm after any number of iterations, even for the complex
energies $z$).

Now we turn to the four-body case. It is convenient to depict
the four-body system by an icon like
\!\!\begin{tabular}{c} \\[-4mm]
\epsfig{file=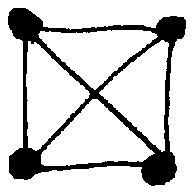,height=1.em}
\end{tabular}\!\!
where the dots denote the particles and the connecting line segments
are associated with both the particle pairs and respective two-body
interactions. There are $n=6$ different pairs (and potentials
$V_\alpha$) that are numbered by $\alpha=1,2,\ldots,6$.

It turns out  \cite{Yakubovsky1967} that the four-body Faddeev
integral equations \eqref{FEint} are not Fredholm even for complex
$z$. This can be understood already from the fact that in the
four-body coordinate space $\mathbb{R}^9$ the supports of any two
finite non-zero two-body potentials $V_\alpha$ and $V_\beta$,
$\beta\neq\alpha$, have an unbounded cylindrical intersection.

Yakubovsky's recipe to tackle the four-body problem consists in the
following. First, introduce the two-cluster partitions:
\!\!\begin{tabular}{c} \\[-4mm]
\epsfig{file=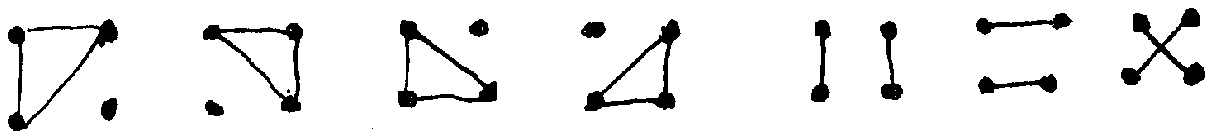,height=1.2em,angle=-1}
\end{tabular}\!\!. Number them by the
Roman letter $a$ (or $b$, $c$ etc.). Then say that a pair $\alpha$
belongs to a partition $a$ and write $\alpha\subset a$ if the
two-body subsystem $\alpha$ belongs to one of the clusters in the
partition~$a$.
For example, \!\!\begin{tabular}{c} \\[-4mm]
\epsfig{file=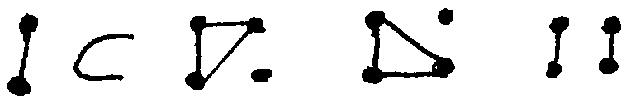,height=0.9em,angle=-1}
\end{tabular}\!\!
while \!\!\begin{tabular}{c} \\[-4mm]
\epsfig{file=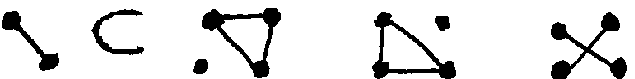,height=0.8em,angle=-0}
\end{tabular}\!\!. In such a case
the sequence $a\alpha$ is called the chain of (consecutive)
partitions. It is assumed that the partition $\alpha$ consists of
one two-particle cluster formed by the pair $\alpha$ and two
one-particle clusters obtained from the partition $a$ by breaking
the corresponding ``bonds''. It is easy to see that there are 18
different chains of partitions: 12 due to $3+1$ and $6$ due $2+2$
starting partitions.

Using the four-body Faddeev integral equations \eqref{FEint}
introduce the following new vectors, called the \textit{Yakubovsky
components} (of the eigenvector $\Psi$):
\begin{equation}
\label{Yakub}
\psi_{a\alpha}=
-(H_0+V_\alpha-z)^{-1}V_\alpha\!\!\sum\limits_{
(\beta\neq\alpha)\subset a}\psi_\beta.
\end{equation}
Then apply to both parts of \eqref{Yakub} the operator
$H_0+V_\alpha-z$ and obtain
\begin{align}
\label{prelast} (H_0+V_\alpha-z)\psi_{a\alpha} &=
-V_\alpha\!\!\!\sum\limits_{(\beta\neq\alpha)\subset
a}\psi_\beta=-V_\alpha\sum\limits_{ (\beta\neq\alpha)\subset
a}\,\,\,\sum_{b\supset\beta}\psi_{b\beta}
\end{align}
taking into account that $\sum_{b\supset\beta}\psi_{\beta
b}=\psi_\beta$. Further, by applying some combinatorics one arrives
at the identity $ \sum\limits_{ (\beta\neq\alpha)\subset
a}\,\,\sum\limits_{b\supset\beta}\psi_{b\beta}
=\sum\limits_{b}\,\,\sum\limits_{ (\beta\neq\alpha)\subset
a}\psi_{b\beta}$ and then from \eqref{prelast} it follows that
\begin{equation}
\label{YE} (H_0+V_\alpha-z)\psi_{a\alpha}+
V_\alpha\sum\limits_{(\beta\neq\alpha)\subset a}\psi_{a\beta}=
-V_\alpha \sum\limits_{b\neq a}\,\,\,
\sum\limits_{(\beta\neq\alpha)\subset a}\psi_{b\beta}.
\end{equation}
These 18 equations for the 18 unknowns $\psi_{a\alpha}$ are just the
desired \textit{Yakubovsky dif\-fe\-ren\-ti\-al equations}.

\begin{center}
\includegraphics[angle=0,width=12.cm]{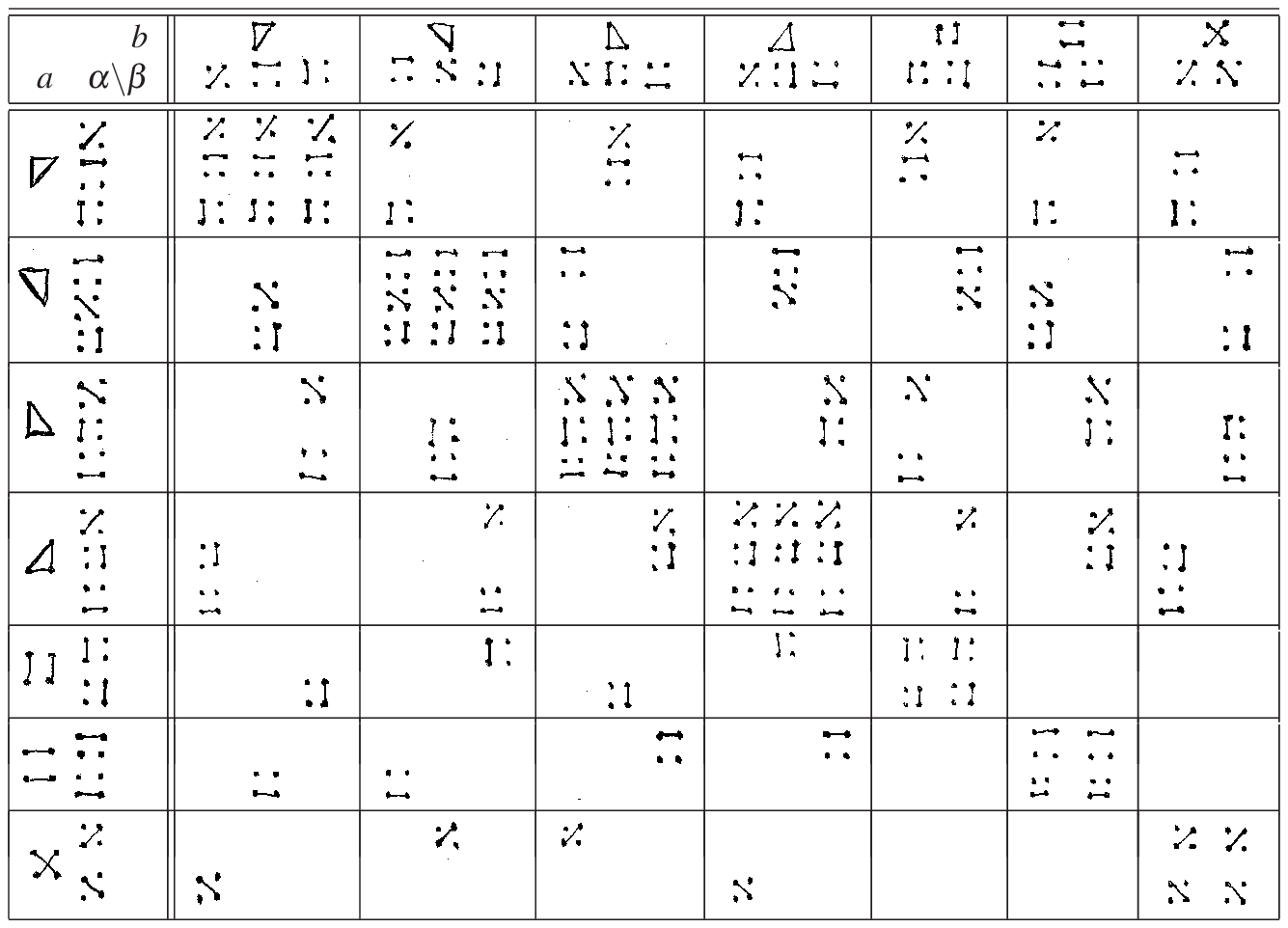}
\end{center}

The table above demonstrates the remarkable structure of the
$18\times18$ block operator matrix associated with the Yakubovsky
equations \eqref{YE}. The icons of three-cluster partitions depict
the indices $\alpha$ of the potentials $V_\alpha$ in the non-zero
entries of the matrix.

\section{Numerical approaches, ideas and tricks}

Compared to the Schr\"odinger equation, the main numerical advantage
of the Faddeev-Yakubovsky differential equations is that the
asymptotical boundary conditions for their physical solutions
imposed at infinitely large distances are much simpler than those
for the total wave function, and they can be much easier
incorporated into the numerical scheme, especially in the case of
scattering processes. Compared to the Faddeev-Yakubovsky integral
equations, it is much easier to tackle the local two-body
potentials. Typically, with such potentials the matrices of the
discretized differential equations have a band structure unlike in
the case of the integral ones.

If particles are identical, the number of the essentially  different
Fad\-de\-ev-Ya\-ku\-bov\-sky components reduces. The components
corresponding to different indices may be obtained from each other
by a simple unitary transformation. For example, the Yakubovsky
equations \eqref{YE} for four identical bosons reduce to a system of
only two equations \cite{MYaG1984}.

The next step is to reduce the dimension of the equations by making
a partial-wave decomposition. Since the total angular momentum $L$
is a conserving quantity, projections onto the reducing subspaces
associated with a fixed value of $L$ keep the emerging equations
exact. At this stage the three Eulerian angles fixing the particles
plane turn out to be separated and the dimension of the Faddeev
equations reduces to 3. This is a modern development to stop at this
stage and solve the three-dimensional Faddeev equations. Such an
approach was first developed in \cite{KKM1989}. It is especially
well suited to the case where two-body interactions depend only on
the distance between particles. The three-dimensional Faddeev
differential equations were first solved numerically in
\cite{HKM1992}.

But historically, the first were the two-dimensional partial-wave
\hbox{(integro-)} differential Faddeev equations where, for a fixed
value of $L$, another variable, the angle between Jacobian vectors,
was eliminated \cite{LG1973}. Further reduction, already to a
one-dimensional form, consists in performing the hyperspherical
adiabatic expansions (see, e.g., \cite{NFJ1998}).

As for the discretization of the equations with respect to the
spatial variables, the finite-difference approximation was initially
used \cite{LG1973}. Starting from \cite{FGPC1984}, for this purpose
various spline approximations are often employed. An interesting
trick with a tensor factorization of the left-hand sides of the
discretized Faddeev equations has been suggested in \cite{SKB1989}.
The factorization trick works very well within iterative approaches
(see \cite{RYaS2005}, \cite{Rou2003}).

Four-body calculations based on the Yakubovsky differential
equations are still rather scarce. Recent examples of such
calculations can be found in \cite{FiYaRV2002,LC2004}.

Latest developments in approaches to solving the Faddeev-Yakubovsky
differential equations are related to the case where the interaction
potentials have an extremely strong repulsive component at small
distances between particles (like, e.g., in atom-atom potentials).
One of the options to tackle such potentials is in approximating the
repulsive part by a hard core. A generalization of the three-body
Faddeev differential equations to the hard-core interactions has
been done yet in 1983 by Merkuriev and the present author
\cite{MM1983}. Differential Yakubovsky equations for this model were
derived in \cite{MMYa1993}. According to \cite{MM1983,MMYa1993}, the
effect of hard cores is reproduced by imposing two-sided boundary
conditions on the Faddeev-Yakubovsky components  at
$x_\alpha=c_\alpha$, $\alpha=1,2,\ldots,n,$ where $x_\alpha$ denotes
the distance between particles of the pair $\alpha$ and $c_\alpha$
the sum of their core radii. In the three-body case these conditions
are quite natural: $\sum_{\beta=1}^3
\psi_\beta\bigr|_{x_\alpha=c_\alpha}=0$,~$\alpha=1,2,3$, while
in the four-body one they are more nontrivial:
\begin{equation}
\label{SummYakubCylinder}
\biggl(\psi_{a\alpha}+\displaystyle\sum_{\beta\subset a}\psi_{a\beta}
+\sum\limits_{b\neq a}\,\,\,\sum\limits_{(\beta\neq\alpha)\subset
a}\psi_{b\beta}\biggr)\biggr|_{x_\alpha=c_\alpha}=0\quad
\text{for any chain } a\alpha.
\end{equation}
It is assumed that $V_\alpha=0$ for $x_\alpha< c_\alpha$ and that
equations \eqref{FEc} and \eqref{YE} are considered on the
respective whole coordinate space $\mathbb{R}^6$ or $\mathbb{R}^9$,
except the hypercylinders $x_\alpha=c_\alpha$ where the hard-core
boundary conditions are imposed. Thus, when a potential with a
strong repulsive core is replaced by the hard-core model, one
approximates inside the core domains only the kinetic energy
operator $H_0$ instead of the sum of $H_0$ and the huge repulsive
term. In this way a much better numerical approximation is achieved.

In recent years this approach was extensively used to calculate
binding energies, resonances, and scattering observables for the
systems of three helium atoms (see papers \cite{MSSK2001,KMS2007}
and references therein). In 2006 the Yakubovsky differential
equations \eqref{YE} with the hard-core boundary conditions
\eqref{SummYakubCylinder} were employed for the first time and with
a big success by La\-za\-us\-kas and Carbonell \cite{LC2006} to
calculate properties of the four-atomic $^4$He$_4$ system. They have
also performed $^4$He$_3$ calculations with the hard-core
Fad\-de\-ev differential equations.

Another (purely numerical) approach to tackle the two-body
potentials with extremely strong repulsion at small distances has
been developed by Ro\-u\-d\-nev and Ya\-kov\-lev. A quite detail
description of this approach is given in \cite{RYaS2005}.

\section{Challenges}

For years the Faddeev-Yakubovsky differential equations showed their
high suitability and efficiency when one solves the three- and
four-body problems, particularly the scattering ones. Recent
advances are related to solving the problems with hard-core (or
practically hard-core) interactions. Still challenging, however, is
the numerical solving of the scattering problems that involve triple
collisions. Also many questions remain unanswered in the case of
few-body scattering problems with Coulomb interactions, especially
with the attractive ones.

\appendix

\section*{\small Appendix: On the spectrum of the Faddeev operator}
\small

Faddeev equations \eqref{FEc} represent the spectral
problem for the $n\times n$ block operator matrix
\begin{equation*}
H_F=\left(\begin{array}{cccc}
H_0+V_1 & V_1 &  \ldots & V_1 \\
V_2 & H_0+V_2 &  \ldots & V_2 \\
\ldots & \ldots & \ldots  & \ldots \\
V_n & V_n &  \ldots & H_0+V_n
\end{array}\right)
\end{equation*}
considered as an operator on the Hilbert space
$\widetilde{\mathfrak{H}}=\overbrace{\mathfrak{H}\oplus\mathfrak{H}
\oplus\ldots\oplus\mathfrak{H}}^{n\text{
times}}$. The matrix $H_F$ is called the \textit{Faddeev operator}
associated with a Hamiltonian of the form
$H=H_0+V_1+V_2+\cdots+V_n$.

The passage from $H$ to $H_F$ adds to $\sigma(H)$ some auxiliary
spectrum. But this auxiliary spectrum is nothing but that of $H_0$.
More precisely, $\sigma(H_F)=\sigma(H)\cup\sigma(H_0)$ (for a proof
see, e.g., \cite{Ya1996}). The spectrum $\sigma(H_0)$ is usually
called the \textit{spurious spectrum} of the Faddeev operator,
although it is quite not ``dangerous'', since it is assumed that one
knows everything about the unperturbed Hamiltonian~$H_0$.

\end{document}